\documentclass[showpacs,preprintnumbers,amsmath,amssymb,preprint]{revtex4}

\usepackage{times}
\usepackage{bm}
\usepackage[dvips]{graphicx}
\usepackage{amsmath}
\usepackage{bm}
\usepackage{natbib}
\usepackage{color}
\usepackage{epstopdf}
\begin{document}
\title{Spin Dynamics in Bilayer Graphene : \\
Role of Electron-Hole Puddles and the Dyakonov-Perel Mechanism}
 \author{Dinh Van Tuan,$^{1}$ Shaffique Adam,$^{2,3}$ and Stephan  Roche$^{1,4}$}
 \affiliation{$^1$Catalan Institute of Nanoscience and Nanotechnology (ICN2), CSIC and The Barcelona Institute of Science and Technology, Campus UAB, Bellaterra, 08193 Barcelona, Spain\\
 $^2$ Centre for Advanced 2D Materials and Physics Department, 
National University of Singapore\\
 $^3$ Yale-NUS College, 138614, Singapore\\
 $^4$ICREA, Pg. Lluís Companys 23, 08010 Barcelona, Spain.}
\date{\today}
\begin{abstract}
We report on spin transport features which are unique to high quality bilayer graphene, in absence of magnetic contaminants and strong intervalley mixing. The time-dependent spin polarization of propagating wavepacket is computed using an efficient quantum transport method. In the limit of vanishing effects of substrate and disorder, the energy-dependence of spin lifetime is similar to monolayer graphene with a M-shape profile and minimum value at the charge neutrality point, but with an electron-hole asymmetry fingerprint. In sharp contrast, the incorporation of substrate-induced electron-hole puddles (characteristics of supported graphene either on ${\rm SiO}_2$ or ${\rm hBN}$) surprisingly results in a large enhancement of the low-energy spin lifetime and a lowering of its high-energy values. Such feature, unique to bilayer, is explained in terms of a reinforced Dyakonov-Perel mechanism at the Dirac point, whereas spin relaxation at higher energies is driven by pure dephasing effects. This suggests further electrostatic control of the spin transport length scales in graphene devices. 
\end{abstract} 

\maketitle

 \textit{Introduction}.- 
Owing to its long spin diffusion length at room-temperature (exceeding several tens of micrometers), single layer graphene (SLG) stands as an unquestionable candidate for the realization of practical devices harvesting the spin degree of freedom, and for more innovation in spintronic applications \cite{Tombros2007,Seneor2012,Dery2012,Han2014,Roche2015}. Such ability to propagate spins over very long distances is due to an intrinsically small spin-orbit coupling (SOC) and hyperfine interaction \cite{Hernando2006}. However, it has been recently demonstrated that the unavoidable coupling between spin and pseudospin degrees of freedom \cite{Rashba2009}, and resulting interwined quantum dynamics, produce a minimum value of spin lifetime ($\tau_{s}$) at the Dirac point, followed by an enhancement of $\tau_{s}$ with energy \cite{Dinh2014,Dinh2015}, as commonly observed experimentally whatever the substrate and material quality \cite{Pi2010,Guimarães2014}.

Bilayer graphene (BLG) differs from SLG by a parabolic band dispersion, however preserving the chiral nature of low-energy electronic excitations. Besides, in contrast to SLG,  an electronic bandgap can be  induced and tuned in BLG under external electric fields \cite{McCann2006}.  Transport measurements show critical differences between SLG and BLG \cite{Bouchiat,Kechedzhi2007,Katsnelson2007}, which are attributed to varying bandstructure and electron-hole puddles characteristics (such as screening strength \cite{Adam2011}). 

Spin lifetimes are also found to differ substantially from the SLG case \cite{BLGTauS}. These differences include variations in the absolute values as well as an opposite scaling of $\tau_{s}$ versus charge density, and a dominating Dyakonov-Perel mechanism in BLG, which lacks a microscopic interpretation \cite{BLGTauS}. Kochan and coworkers \cite{Kochan2014} have argued that magnetic contaminants are necessary to explain the energy-dependent profiles and short values of spin lifetimes in monolayer and bilayer graphene,  but their approach also supposes an Elliot-Yafet spin relaxation mechanism in bilayer, which is contradicted by experiments \cite{BLGTauS,Soriano2015}.

In this Letter, thanks to a fully quantum treatment of spin dynamics in real space, we report on spin transport features which are unique to high quality BLG. The critical role of the substrate to reproduce the typical energy-dependent profile of spin lifetimes (observed experimentally) is revealed, with a large enhancement of $\tau_{s}$ near the charge neutrality point driven by electron-hole puddles and local electric field effects. By comparison with the unsupported (pristine) BLG, our findings identify that a reinforcement of the Dyakonov-Perel mechanism occurs at low energy, whereas higher energy spin lifetimes are dictated by quantum dephasing effects.  Our study points towards the uniqueness of the BLG bandstructure in presence of spin-orbit interaction and non-uniform energy-dependent spin precession frequency to capture spin transport fingerprints, and suggest the possibility to electrostatically monitor the spin diffusion length scales.

\begin{figure}[htbp]
 \includegraphics[width=0.6\textwidth]{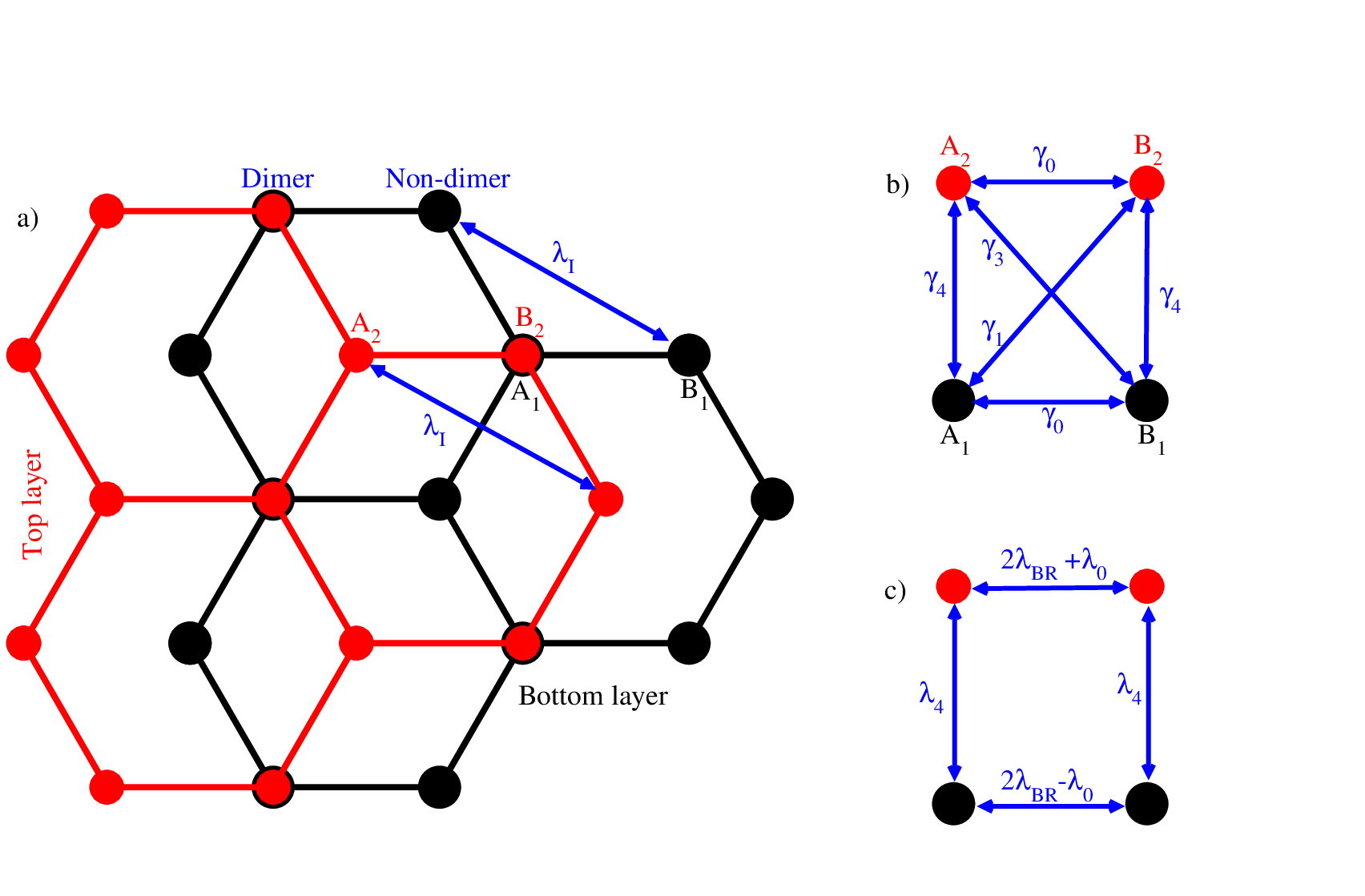}
\caption{(color online): (a) Sketch of BLG composed of a top (in red) and a bottom (in black) layers. The  intralayer intrinsic SOC is shown in blue. (b) Schema of non-spin-orbit interaction between one type of carbon atom
to the nearest carbons of the other kind. (c) The same in (b) but for Rashba-type SOC.}
 \label{Fig1}
 \end{figure}

\textit{BLG Hamiltonian in presence of spin-orbit interaction and electron-hole puddles}.- BLG  can be considered as two coupled  SLGs with the top layer shifted a carbon bond from the bottom layer (Fig.\ref{Fig1}(a)). Consequently, BLG consists of four carbon atoms in its unit cell, two carbons  $A_1,B_1$ from the unit cell of the bottom SLG and $A_2, B_2$ from the top layer where $B_2$ places  on the top of $A_1$, namely dimer sites and $B_1, A_2$ are called non-dimer sites.  In the tight-binding model the full Hamiltonian for BLG reads :
\begin{equation}
{\mathcal{H}}={\mathcal{H}_{SLG}^T}+{\mathcal{H}_{SLG}^B}+{\mathcal{H}^{Inter}_0}+\mathcal{H}_{SOC}^{Inter} 
\label{FullHalBLG}
\end{equation}
where the first and second terms (intralayer parts ${\mathcal{H}_{SLG}^l}$) are the  Hamiltonians for each single layer involving the SOCs effect (intrinsic $\lambda_I$ and Rashba types $\lambda_R^l$), the different potential energies $\Delta$ of the top ($l=1$) and bottom layer ($l=2$) as well as the long range potential simulating the electron-hole puddles $V(r)$ \cite{Adam2011}. 
 \begin{eqnarray}
{\mathcal{H}_{SLG}^l}=&-&\gamma_0\sum_{\langle ij\rangle }a_{l,i}^+b_{l,j}+\frac{2i}{3}\sum_{\langle ij\rangle}\lambda_R^l a_{l,i}^+\vec{z}\cdot(\vec{s}\times\vec{d}_{ij})b_{l,j}\nonumber\\
&+&\frac{2i}{9}\lambda_{I}\sum_{\langle\langle ij\rangle\rangle}a_{l,i}^+\vec{s}\cdot(\vec{d}_{kj}\times\vec{d}_{ik})b_{l,j} +h.c.\nonumber\\
&-&\frac{\Delta}{2}\sum_{ i }(-1)^l\left\{a_{l,i}^+a_{l,i}+b_{l,i}^+b_{l,i}\right\} +V(r)
\label{HamilSLG}
\end{eqnarray}
Where $a_{l,i}$ ($b_{l,i}$) is the annihilation operators acting on $A_i$ ($B_i$) in layer $l$.
 $\vec{d}_{ij}$ is the unit vector pointing from $j$ to $i$, $k$ is the common nearest neighbor of $i$ and $j$.
 In this paper we use intralayer-intrinsic SOC  $\lambda_I=12\mu$ eV \cite{Konschuh2012} (Fig.\ref{Fig1}(a)), and intralayer-Rashba SOC $\lambda_R^l=2\lambda_{BR}- (-1)^l\lambda_0$ (Fig. \ref{Fig1}(c)) which involves two contributions, one from the bulk-inversion-asymmetry
 induced by the adjacent layer with $\lambda_0=5\mu {\rm eV} $ \cite{Konschuh2012} and  the other $2\lambda_{BR}=10\times E [{\rm V/nm}] \mu {\rm eV}$  which is field dependent. In this calculation we choose $2\lambda_{BR}=2.5 \mu$eV corresponding to an electric field $E=0.25 [{\rm V/nm}]$  independent of the charge density \cite{Konschuh2012}, which is a reasonable approximation from experimental considerations \cite{Avsar2015}. The third term in Eq.(\ref{FullHalBLG}) is the non-spin-orbit coupling part of the interlayer Hamiltonian 
 \begin{eqnarray}
{\mathcal{H}_0^{Inter}}&=&\gamma_1\sum_{ i }a_{1,i}^+b_{2,i}
+\gamma_3\sum_{ \langle ij\rangle }b_{1,i}^+a_{2,j}\nonumber\\
&-&\gamma_4\sum_{ \langle ij\rangle }\left\{a_{1,i}^+a_{2,j}+b_{1,i}^+b_{2,j}\right\}+h.c. \nonumber
\end{eqnarray}
where the first term in above Hamiltonian describes the interlayer hopping ($\gamma_1=340$ meV) between dimer sites $\{A_1,B_2\}$ \cite{Konschuh2012}. The second term denotes the interlayer coupling between $B_1$ and its adjacent $A_2$ with $\gamma_3=280$ meV. The third term corresponds to hopping integral from $A_1$ to its adjacent $A_2$, and from $B_1$ to its adjacent $B_2$, with $\gamma_4=145$ meV (Fig. \ref{Fig1}(b)). All these parameters have been derived from the {\it ab-initio} calculations \cite{Konschuh2012,Jung2014}. Finally, the SOC part of the interlayer interaction is described by the final term in Eq.(\ref{FullHalBLG}), ${\mathcal{H}_{SOC}^{Inter}}$ which reads :

$$
\frac{2i\lambda_4}{3}\sum_{i,j}\left\{ a_{1,i}^+\vec{z}\cdot(\vec{s}\times\vec{d}_{ij}^\parallel)a_{2,j}+b_{1,i}^+
\vec{z}\cdot(\vec{s}\times\vec{d}_{ij}^\parallel)b_{2,j}\right\}+h.c.
$$

where $\vec{d}_{ij}^\parallel$ is the unit vector of the projection of vector $\vec{d}_{ij}$ on the horizontal plane. This part is the Rashba-type spin-orbit interaction ($\lambda_4=-12\mu {\rm eV}$ \cite{Konschuh2012}) between sites with interlayer hopping term described by $\gamma_4$ (Fig. \ref{Fig1}(c)). Additional much smaller terms defined in Ref. \cite{Konschuh2012} are found to bring a negligible contribution to the results.

The electron-hole puddles in BLG are simulated by a long-range potential $V({\bf r})=\sum_{j=1}^{N}\epsilon_j\exp[-({\bf r}-{\bf R}_j)^2/(2\xi^2)] $, with $\xi=3.5$ nm, a value extracted from self-consistent calculations \cite{Adam2011}. The magnitude of the onsite potential $\epsilon_j$ is randomly chosen within $\left[-W,W\right]$ with $W=35$ meV for ${\rm SiO}_2$ substrate and $W=11 $ meV for hBN substrate \cite{Adam2011}.
 The impurity concentration is $n_i=10^{12}{\rm cm}^{-2} (0.04\%)$ for a ${\rm SiO}_2$ substrate and $n_i=10^{11}{\rm cm}^{-2} (0.004\%)$ for a hBN substrate.
 
\vspace*{7mm}
  
\textit{Spin dynamics methodology.}- The spin dynamics of electron in BLG is investigated using the time-dependent evolution of the spin polarization ${P_z}(E,t)$ of propagating wavepackets \cite{Dinh2014}, which is computed through
\begin{equation}
{P_z}(E,t)=\frac{\langle\Psi(t)| s_z\delta(E-{\mathcal{H}})+\delta(E-{\mathcal{H}})s_z|\Psi(t)\rangle}{2\langle\Psi(t)|\delta(E-{\mathcal{H}})|\Psi(t)\rangle}
\label{time-dependence0}
\end{equation}
where $s_z$ is the $z$ component of the Pauli matrices and $\delta(E-{\mathcal{H}})$ is the spectral measure operator. The evolution of the wavepackets $|\Psi(t)\rangle$ is obtained by solving the time-dependent Schr\"{o}dinger equation \cite{Roche2014b}, starting from a wavepacket $|\Psi(t=0)\rangle$ in an out-of-plane ($z$ direction)  polarization. An energy broadening parameter $\eta=13.5$ meV is introduced for expanding $\delta(E-{\mathcal{H}})$ through a continued fraction expansion of the Green's function \cite{Roche2014b}. This method has been previously used to investigate spin relaxation in gold-decorated graphene \cite{Dinh2014}, hydrogenated graphene \cite{Soriano2015}, or SOC coupled graphene under the effect of electron-hole puddles \cite{Dinh2015}.  

\begin{figure}[htbp]
\includegraphics[width=0.5\textwidth]{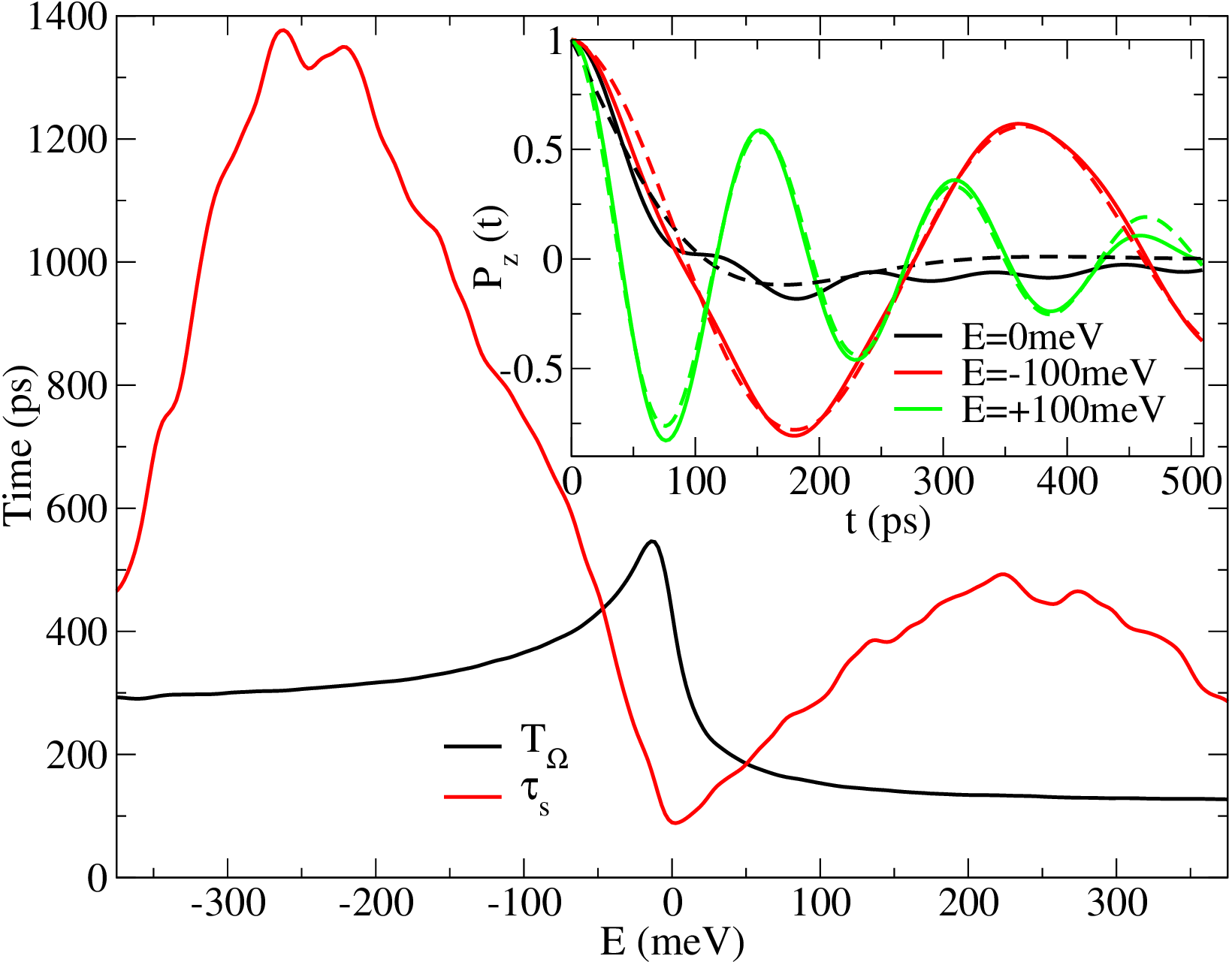}
\caption{(color online) Main frame: Spin lifetime $\tau_s$ (red line) and spin precession time $T_\Omega$ (black line) in spin-orbit coupled BLG, in absence of electron-hole puddles. In this limit, $\tau_s$ has the same characteristic M-shape as for monolayer graphene, but with some electron-hole asymmetry due to skew interlayer hopping. Inset: spin polarization $P_z(t)$ at some energies (solid lines) together with fits to the function $P_z(t)=\cos (2\pi t/T_\Omega)e^{-t/\tau_s}$ (dashed lines) from which $T_\Omega$ and $\tau_s$ are extracted (main frame). }    
\label{Fig2}
\end{figure}

\textit{Spin dynamics and dephasing in the ultraclean limit}- We first consider the situation of pristine BLG in absence of microscopic disorder ($V(r)=0$; $\Delta=0$), and in which only the uniform SOC and the small energy broadening dictate the spin lifetime characteristics. Fig. \ref{Fig2} (inset) shows $P_z(E,t)$ (solid lines), which exhibits an oscillatory pattern typical for spin precession together with an exponential decay which dictates the loss of spin information. A significant electron-hole asymmetry is observed for two chosen energies $E=\pm 100$ meV, which correspond to a charge density of $\pm 5\times 10^{12}{\rm cm}^{-2}$. A faster oscillation of the spin signal is seen at $E=100$ meV (green line) in contrast to the slower oscillation observed at $E=-100$ meV (red line).  The spin relaxation at the charge neutrality point is even faster and likely driven by the same interwoven dynamics of spin and pseudospin degrees of feeedom, as unveiled for SLG \cite{Dinh2014}.

From the fits of the numerical data using $P_z(t)=\cos (2\pi t/T_\Omega)\exp (-t/\tau_s)$ (dashed lines), the spin precession time $T_\Omega$ and $\tau_s$ are extracted (see Fig. \ref{Fig2} (inset), dashed lines). The electron-hole asymmetry is observed in both $T_\Omega$ and $\tau_s$ with larger values for the hole side compared to the electron side.  This phenomenon has been noticed by Diez and Burkard \cite{Diez2012}, who proved that the contribution of skew interlayer hopping term $\gamma_3$ leads to the reduction (increase) of spin splitting energy $\Delta E$ in the hole (electron) side (note that $T_\Omega\sim 1/\Delta E$). The largest variation of $T_\Omega (E)$ occurs in the vicinity of the charge neutrality point, and this non-uniformity results in stronger dephasing effects and shortest spin lifetime, as discussed for the SLG case \cite{Dinh2014,Dinh2015}. On the other hand, the obtained value for $\tau_s$ varies from $100$ ps to $1.4$ ns, which is the typical range of experimental data \cite{BLGTauS}.  Finally,  $\tau_s(E)$ for BLG shares another similar additional feature with SLG, that is a downturn at higher energies, which has been related to the contribution of trigonal warping \cite{Dinh2015}.   

\vspace*{7mm}
 \begin{figure}[htbp]
 \includegraphics[width=0.5\textwidth]{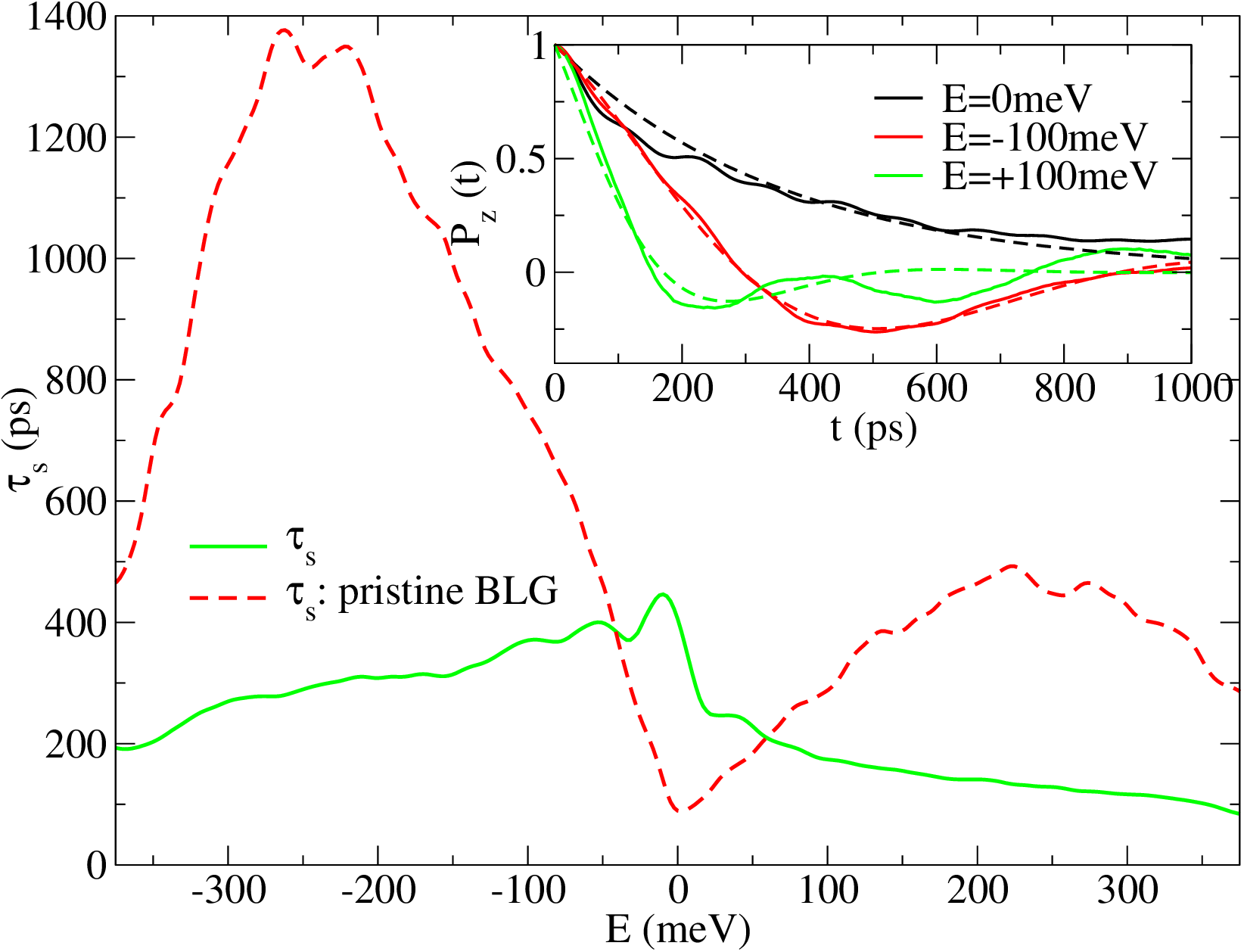}
\caption{(color online) Main frame: spin lifetime $\tau_s$ (solid line) for graphene on ${\rm SiO}_2$ substrate in the comparison with $\tau_s$ of pristine spin-orbit coupled BLG (dashed line). Inset: spin polarization $P_z(t)$ at some energies (solid lines) and their  fits to the function  $P_z(t)=\cos (2\pi t/T_\Omega)e^{-t/\tau_s}$ (dashed lines)}
 \label{Fig3}
 \end{figure}

To study the substrate effect, we introduce electron-hole puddles through the long range potential $V(r)$, which well reproduce the measured charge density fluctuations for graphene either supported on ${\rm SiO}_2$ or hBN \cite{Adam2011}. Fig. \ref{Fig3} (inset) shows the evolution of spin polarization $P_z(t)$ (solid lines) for graphene on ${\rm SiO}_2$ substrate and the corresponding fits (dashed lines) from which  $\tau_s$ is extracted (main frame, green solid line). Remarkably, the electron-hole puddles associated to ${\rm SiO}_2$ substrate provokes an inversion in the energy-dependent profile of $\tau_s$, with a peak at CNP, as reported in experiments in the low temperature regime \cite{BLGTauS}. The absolute values, energy dependence as well as the electron-hole  asymmetry of the extracted $\tau_s$ provide a consistent support to the analysis of state-of-the-art experimental data without the need to introduce any magnetism in the problem \cite{BLGTauS}. It is worth mentioning that the obtained value for $\tau_s$  (green solid line) close to CNP is larger than in the case of pristine spin-orbit coupled BLG with no puddles (red dashed line). Such enhancement of $\tau_s$ close to CNP, and driven by electron-hole puddles, can be rationalized as a reinforcement of the Dyakonov-Perel mechanism \cite{Diez2012}. Indeed, the presence of electron-hole puddles generates elastic scattering which act on the spin precession and produce a motional narrowing phenomenon. This is particularly strong at the charge neutrality owing to the specific bandstructure of BLG where parabolic bands and higher density of states are obtained and favour an enhancement of the scattering probability, when compared to SLG \cite{Nilsson2008}. This enhanced scattering more efficiently impedes spin precession, which in absence of disorder, is the mechanism driving to relaxation. Such effect is inactive in SLG \cite{Dinh2015}, bringing an essential difference between both types of structures. 

\begin{figure}[htbp]
 \includegraphics[width=0.5\textwidth]{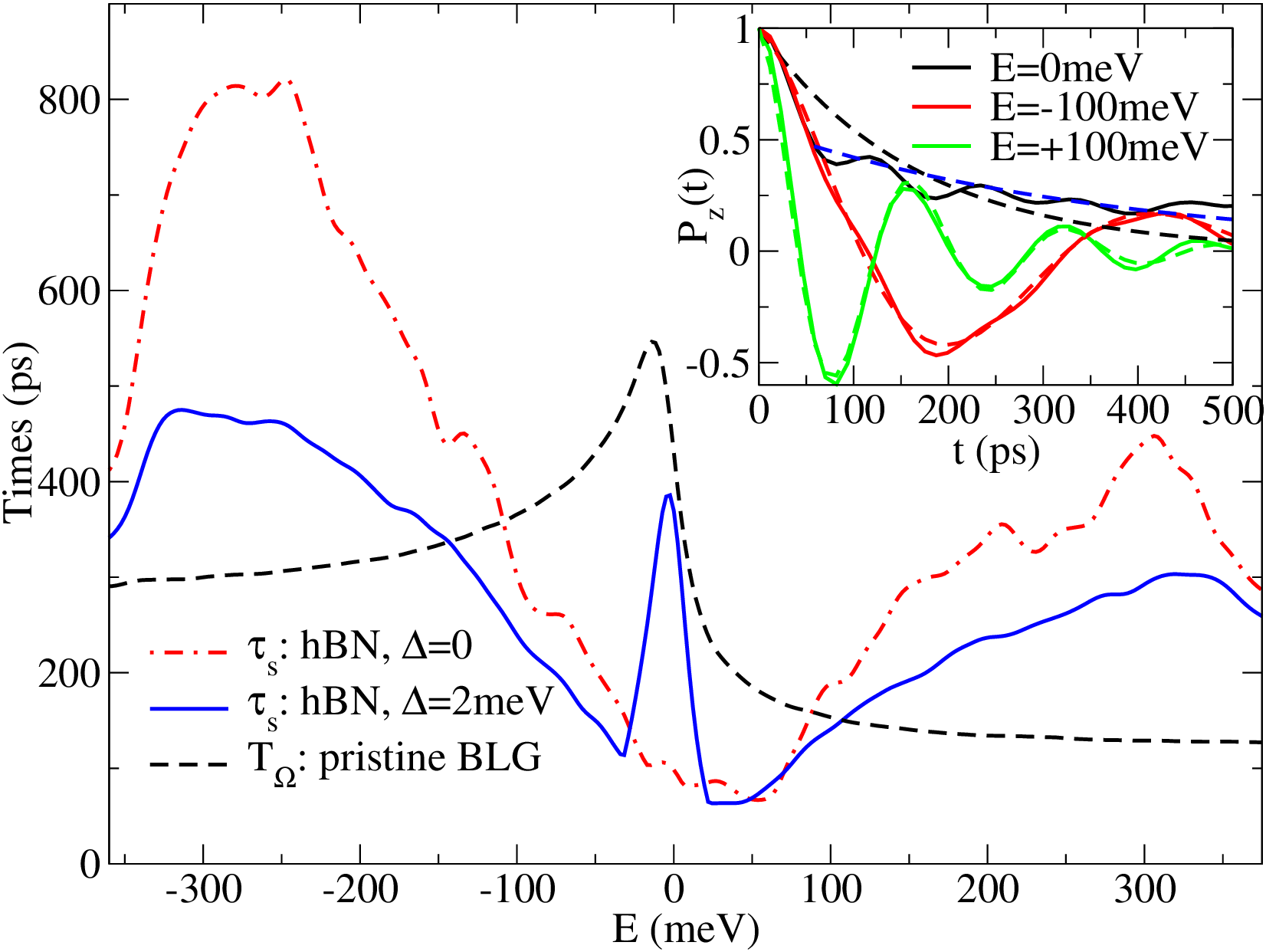}
\caption{(color online):  Main frame: Spin relaxation time $\tau_s$ for graphene on $hBN$ substrate (i.e. with the corresponding electron-hole puddles), with (blue line) or without (dotdashed red line) the additional energy asymmetry between top and bottom layers. $T_\Omega$ for the pristine spin-orbit coupled BLG is shown in black dashed line. Inset: spin polarization $P_z(t)$ at some energies (solid lines) together with fits to $P_z(t)=\cos (2\pi t/T_\Omega)e^{-t/\tau_s}$ (dashed lines).  For states close to CNP, the fit is made with $P_{z}(t)=P_{z}(t_0)e^{-(t-t_0)/\tau_s}$.}
 \label{Fig4}
 \end{figure}

For the case of hBN substrate, the situation becomes more complicated. Scattering due to electron-hole puddle alone ($W=11$ meV and $n_i=10^{11}{\rm cm}^{-2}$) is too weak to modify the results when compared to the unsupported case (Fig.\ref{Fig1}). However, the scattering strength close to CNP due to electron-hole puddles is likely enhanced by the formation of a pseudogap induced by the electric field, so far neglected. Indeed in presence of an external electric field, the weak interaction between hBN and graphene layers breaks the symmetry between top and the bottom layers, an effect which can be modelled by adding an energy difference between layers \cite{Ramasubramaniam2011,McCann2006}. To account for it, we thus introduce a small energy difference between the top and the bottom layers, which would open a real gap of $\Delta=2$ meV for the pristine BLG. Here however, the disorder potential stemming from the electron-hole puddles is strong enough to wipe out the gap ($W\gg \Delta$), while maintaining spatially uncorrelated local energy fluctuations between layers.

Fig. \ref{Fig4} shows the resulting spin lifetime for BLG on hBN. One first observes that by setting $\Delta=0$, the spin lifetime decreases (dotdashed red line) when compared to the unsupported clean case (see Fig. 2), while keeping a similar shape over the whole spectrum. In sharp contrast, the presence of a small extra energy asymmetry between top and bottom layers ($\Delta\neq 0$) results in a substantial enhancement of multiple scattering effects at low energy. Similarly to the case of ${\rm SiO}_2$ substrate, the Dyakonov-Perel mechanism is then reinforced close to the CNP and spin relaxation is reduced. For $\Delta=2$ meV, the time dependence of spin polarization close to CNP changes dramatically from the behavior $P_z(t)=\cos (2\pi t/T_\Omega)e^{-t/\tau_s}$ to an exponential decay $P_{z}(t)=P_{z}(t_0)e^{-(t-t_0)/\tau_s}$ (fit starts from $t_0=60$ ps, see blue dashed line in inset), which clearly shows that the spin precession is more strongly suppressed.

The extracted $\tau_s$ is seen to be very sharp close to CNP (as observed in a recent experiment \cite{Avsar2015}). Such enhancement of $\tau_s$ originates from the more efficient motional narrowing driven by the Dyakonov-Perel mechanism. Indeed, the strong variation of spin precession time $T_\Omega$ (black dashed line) close to CNP gives rise to spin dephasing and short $\tau_s$. Here, for Gaussian correlated disorder with the chosen parameters, a semi-classical transport calculation gives a minimum momentum scattering time $\tau_p = \frac{1}{4\pi} \left(\frac{\hbar^2}{m \xi^2 W} \right) \frac{\hbar}{W} \approx 100~{\mbox fs}$ for a hBN substrate, and about $10~{\mbox fs}$ for bilayer graphene on a SiO$_2$ substrate.  We note that experimentally $\tau_p$ for BLG on SiO$_2$ \cite{Avsar2015,Monteverde2010} and on hBN \cite{Ramasubramaniam2011} are consistent with these estimates, and in all cases $\tau_p \ll T_\Omega$ which satisfies the criterion to enter into the DP regime and supports our interpretation of an enhanced DP mechanism at the Dirac point. Overall, scattering events induced by electron-hole puddle together with the pseudogap act against the spin dephasing of cleaner samples and consequenly leads to the enhancement of spin relaxation time close to CNP, indicating that the Dyakonov-Perel mechanism governs the low-energy spin lifetime in BLG. 

Finally, It is worth mentioning that the pseudogap of graphene on hBN is not only induced by electric field but also by the staggered potential which captures the interaction between the graphene lattice and hBN \cite{Jung2015}. We estimate that for hBN, the moir\'{e} band approximation is relevant given the low level of disorder produced by electron-hole puddles \cite{DaSilva2015}. This can explain the fact that we observe a sharp peak for spin relaxation time on hBN whereas a broaden peak is experimentally observed for graphene on ${\rm SiO}_2$.

{\it Discussion and conclusion.}- Recently, D. Kochan and coworkers \cite{Kochan2014} suggested that the origin for spin relaxation in BLG is the same with SLG, namely resonant scattering by magnetic impurities. By varying the strength of electron-hole puddles and broadening factors, the experimental data could roughly reproduce $\tau_s$ with the upturn at CNP in BLG or the downturn in SLG. However, this scenario predicts an Elliot-Yafet type of relaxation \cite{Soriano2015}, whereas experiments on BLG clearly evidence a scaling behavior as $\tau_s\sim 1/\tau_{p}$, indicating the predominance of the Dyakonov-Perel mechanism \cite{BLGTauS}. Here we have found that even for long mean free paths, the impact of electron-hole puddles due to silicon oxide or hBN substrates make the Dyakonov-Perel mechanism predominating over spin dephasing, a fact unique to BLG. A characteristic peak at the CNP will be seen for the ${\rm SiO}_2$ or will be very sharp around the CNP for the hBN substrate. Those findings provide a consistent interpretation of all reported experiments on BLG, without the need of introducing additional relaxation mechanism driven by magnetic impurities \cite{BLGTauS,Avsar2015}.  

This project has received funding from the European Union's Horizon 2020 research and innovation programme under grant agreement No 696656, Graphene Flagship. S.R. acknowledges Funding from the Spanish Ministry of Economy and Competitiveness and the European Regional Development Fund (Project No. FIS2015-67767-P (MINECO/FEDER)) and the Severo Ochoa Program (MINECO SEV-2013-0295). S.A. acknowledges support from the National Research Foundation of Singapore under its Fellowship  programme  (NRF-NRFF2012-01) and computational resources from the Centre for Advanced 2D Materials (an NRF mid-sized centre).

\end{document}